\documentclass[preprint]{aastex61}

\usepackage{amsmath}
\usepackage{amssymb}
\usepackage{graphicx}
\usepackage{bm}
\usepackage{geometry}
\usepackage{natbib}
\usepackage{extarrows}
\usepackage{color}
\usepackage{ulem}

\begin{document}

\title{Two-zone diffusion of electrons and positrons from Geminga 
explains the positron anomaly}

\author{Kun Fang}
\affiliation{Key Laboratory of Particle Astrophysics, Institute of High Energy 
Physics, Chinese Academy of Sciences, Beijing 100049, China}
\affiliation{School of Physics, University of Chinese Academy of Sciences, 
Beijing 100049, China} 

\author{Xiao-Jun Bi}
\affiliation{Key Laboratory of Particle Astrophysics, Institute of High Energy 
Physics, Chinese Academy of Sciences, Beijing 100049, China}
\affiliation{School of Physics, University of Chinese Academy of Sciences, 
Beijing 100049, China} 

\author{Peng-Fei Yin}
\affiliation{Key Laboratory of Particle Astrophysics, Institute of High Energy 
Physics, Chinese Academy of Sciences, Beijing 100049, China}
 
\author{Qiang Yuan}
\affiliation{Key Laboratory of Dark Matter and Space Astronomy, Purple Mountain 
Observatory, Chinese Academy of Sciences, Nanjing 210008, China}
\affiliation{School of Astronomy and Space Science, University of Science and 
Technology of China, Hefei, Anhui 230026, China}

\date{\today}

\begin{abstract}
The recent HAWC observations of very-high-energy $\gamma$-ray halo
around Geminga and Monogem indicate a very slow diffusion
of cosmic rays， which results in tiny contribution of positrons from these two 
pulsars to the local flux. This makes the
cosmic positron excess anomaly observed by PAMELA and AMS-02 even more puzzling.
However, from the Boron-to-Carbon ratio data one can infer that the
average diffusion coefficient in the Galaxy should be much larger.
In this work we propose a two-zone diffusion model that the diffusion 
is slow only in a small region around the source, outside of which
the propagation is as fast as usual. We find that such a scenario can 
naturally explain the positron excess data with parameters even more
reasonable than that in the conventional one-zone diffusion model.
The reason is that during the life time of Geminga ($\sim 300$ kyr) the 
electrons/positrons have propagated too
far away with a fast diffusion and lead to a low local flux. 
The slow diffusion region in the two-zone model helps to confine
the electrons/positrons for a long time and 
lead to an enhancement of the local flux. So under the constraint of the HAWC 
observations, pulsars are still the probable origin of the cosmic-ray positron 
excess.
\end{abstract}

\section{Introduction}
\label{sec:intro}
The cosmic-ray positron excess has been discovered for nearly a decade 
\citep{2009Natur.458..607A,2012PhRvL.108a1103A,2013PhRvL.110n1102A},
but its origin is still a mystery. These extra positrons may originate 
either from astrophysical sources like pulsars or the dark matter 
annihilation/decay. Among various kinds of astrophysical sources, the 
Geminga pulsar (PSR J0633+1746) has been widely believed to be a very 
promising candidate to produce the positron excess 
\citep{2009PhRvL.103e1101Y,2009JCAP...01..025H,2013PhRvD..88b3001Y,
2017PhRvD..96j3013H}. Geminga is one of the nearest pulsars with a 
distance of $250^{+120}_{-62}$ pc \citep{2007Ap&SS.308..225F}. 
Its age is estimated to be about $3.42\times10^5$ years, and the 
derived spin-down energy is $1.23\times10^{49}$ erg 
\citep{2005AJ....129.1993M,2018ApJ...854...57F}. All these parameters 
suggest that Geminga can probably dominate the high energy positron 
flux observed on the Earth. The extended TeV $\gamma$-ray halo around 
Geminga pulsar observed by Milagro \citep{2007ApJ...664L..91A} and HAWC 
\citep{2017ApJ...843...40A} gives straightforward evidence supporting
that Geminga can indeed generate very high energy electrons and 
positrons ($e^\pm$). 

However, the detailed morphological study of the very high-energy 
$\gamma$-ray emission from Geminga and PSR B0656+14 (Monogem) by
HAWC suggests that the $e^\pm$ produced by these pulsars diffuse out
significantly slower than that in the average interstellar medium (ISM)
as inferred from the Boron-to-Carbon ratio (B/C) measurements
\citep{2017Sci...358..911A}. In such a case, the $e^\pm$ produced by 
Geminga or Monogem can hardly reach the Earth, and thus these two 
pulsars may be unlikely to account for the positron excess. 
On the other hand, \citet{2017arXiv171107482H} pointed out that this 
slow-diffusion scenario should not be representative even in the local 
environment. Since H.E.S.S. has detected high-energy $e^\pm$ up to 
$\sim$ 20 TeV\footnote{
https://indico.snu.ac.kr/indico/event/15/session/5/contribution/694},
these $e^\pm$ can only travel for $10\sim20$ pc within the cooling time
given such a slow-diffusion condition. We can hardly find any high-energy 
$e^\pm$ sources within such a small distance around the solar system.

Since the HAWC data can only probe a region of $\sim30$ pc around
Geminga and Monogem, it is very likely that the slow-diffusion region
is actually limited in a small region around the sources, beyond
which particles diffuse faster as typical Galactic cosmic rays.
This scenario would be consistent with the B/C data and the H.E.S.S 
$e^\pm$ spectrum \citep{2017PhRvD..96j3013H,2017arXiv171107482H}. 
In this work, we investigate whether this two-zone diffusion model can 
explain both the HAWC $\gamma$-ray data of Geminga and the positron 
excess. The two-zone diffusion model is solved with a numerical method.
Because the diffusion coefficient has a jump at the boundary between 
the two zones, the differencing scheme should be carefully dealt with.

In Section \ref{sec:method}, we introduce the two-zone diffusion model and 
the numerical treatment to the propagation equation. In Section \ref{sec:rst}, 
we calculate the positron spectrum of Geminga in the two-zone scenario, and 
compare the result with the AMS-02 data \citep{2014PhRvL.113l1102A}. Then we 
conclude our work with some discussion in the last section.

\section{Method}
\label{sec:method}
The propagation process of $e^\pm$ can be described by the diffusion-cooling 
equation 
\begin{equation}
 \frac{\partial N}{\partial t} - \nabla(D\nabla N) - \frac{\partial}{\partial
E}(bN) = Q \,,
 \label{eq:prop}
\end{equation}
where $N$ is the differential number density of $e^\pm$, $D$ denotes the
diffusion coefficient, $b$ is the energy-loss rate, and $Q$ is the source 
term. In the present work, we are interested in the energy range higher 
than 10 GeV, so the convection and reacceleration terms which affect the
low energy spectrum are neglected \citep{dela09}. The energy-loss rate has the 
form of $b(E)=b_0(E)E^2$, which describes the synchrotron and inverse Compton 
radiation cooling of $e^\pm$. The interstellar magnetic field in the Galaxy is 
set to be 3 $\mu$G to get the synchrotron term \citep{1996ApJ...458..194M}. 
For the inverse Compton scattering term, it is necessary to consider a 
relativistic correction to the scattering cross section. We follow the 
calculation of \citet{schli10}, where $b_0$ is energy-dependent.

The diffusion coefficient is usually assumed to be $D(E)=\beta^\eta 
D_0{(R/\rm 1\,GV)}^{\delta}$, where $D_0$ and $\delta$ are both constants, 
$\beta$ is the velocity of particles in unit of light speed, $\eta$ is
a low energy correction parameter of the velocity dependence, and $R$ is 
the rigidity. Assuming $D$ is spatially constant, \citet{2017PhRvD..95h3007Y} 
constrained the propagation parameters with the B/C data of AMS-02 
\citep{2016PhRvL.117w1102A}, and found that the best-fit model is the 
diffusion model with reacceleration. The obtained propagation parameters
are $D_0=(2.08\pm0.28)\times10^{28}$ cm$^2$ s$^{-1}$ and 
$\delta=0.500\pm0.012$ (hereafter Y17 model). The thickness of the propagation 
halo is $z_h=5.02\pm0.86$ kpc, and the low energy correction parameter $\eta$ 
is not relevant for this study. 

For the $e^\pm$ produced by Geminga, the propagation distance can be 
estimated as $2\sqrt{D(E)t_E}$, where $t_E={\rm min}\{t_{\rm g},1/(b_0E)\}$, 
i.e., the smaller one of the age of Geminga $t_{\rm g}$ and the cooling 
time \citep{2017Sci...358..911A}. If we adopt the propagation parameters 
of Y17, $e^\pm$ of $\simeq 1$ TeV energies can diffuse to a distance of 
$\simeq 1.7$ kpc. This scale is much smaller than $z_h$ obtained in Y17. 
If the particles diffuse slower in the region around Geminga, the
total diffusion distance is even less. Therefore, it should be fine
to assume a spherically symmetrical geometry of the propagation of
$e^\pm$ from Geminga. The diffusion coefficient for the two-zone model 
is then
\begin{equation}
 D(E, r)=\left\{
 \begin{aligned}
  D_1(E), & & r< r_\star \\
  D_2(E), & & r\geq r_\star\\
 \end{aligned}
 \right.\,,
 \label{eq:diff}
\end{equation}
where $r$ is the distance from Geminga, $r_\star$ is the discontinuity 
shell of the diffusion coefficient, $D_1$ is the diffusion coefficient 
around Geminga which is inferred by the HAWC $\gamma$-ray data
\citep{2017Sci...358..911A}, and $D_2$ is the average diffusion
coefficient in the Milky Way given by Y17. In a more realistic picture the 
diffusion coefficient may change gradually. However, no observations can 
constrain the transitional zone. Further as $r_\star$ is a free parameter we 
can always get a equivalent result for the local $e^\pm$ flux by adjusting 
$r_\star$ in this simplified picture. 

The spatially-dependent diffusion equation is difficult to be solved 
analytically, and we adopt a numerical method instead in this work.
We assume that Geminga is a burst-like point source, then Equation 
(\ref{eq:prop}) can be rewritten as following, along with the initial condition 
and the boundary conditions:
\begin{equation}
 \left\{
 \begin{aligned}
  & \frac{\partial N}{\partial t} = \mathcal{L}N\,, \\
  & N(0, E, r)=Q(E)\delta(r)\,, \\
  & N(t, E_{\rm max}, r)=0\,, \\
  & \left.\frac{\partial N}{\partial r}\right|_{r=0}=0\,, \\
  & N(t, E, r_{\rm max})=0\,, \\
 \end{aligned}
 \right.
 \label{eq:prop_set}
\end{equation}
where $E_{\rm max}=500$ TeV, and $r_{\rm max}=4$ kpc. The operator 
$\mathcal{L}$ is the sum of the diffusion operator $\mathcal{L}_r$ and the 
energy-loss operator $\mathcal{L}_E$, which are
\begin{equation}
 \begin{aligned}
  & \mathcal{L}_r=\frac{1}{r^2}\frac{\partial}{\partial 
r}\left[r^2D(r)\frac{\partial}{\partial r}\right]\,, \\
  & \mathcal{L}_E= b\frac{\partial}{\partial E}+\frac{\partial b}
{\partial E}\,. \\
 \end{aligned}
 \label{eq:operator}
\end{equation}
We apply the {\it operator splitting method} to deal with the two operators 
separately. For the discretization of the energy-loss scheme, we follow the 
method given by \citet{2014APh....55...37K}, while for the diffusion 
operator, we use two different discretizations depending on if $r_i$ 
lies on the discontinuity surface ($i$ is the spatial step index). If not, the 
well-known Crank-Nicolson scheme for constant $D$ is a good choice. For the 
case of $r_i=r_\star$, the difference equation should be re-written to ensure 
the conservation of the flux at both sides of the discontinuity surface. We 
adopt the {\it finite volume method} to derive the differencing scheme for 
this case. All the details of the discretization are presented in Appendix 
\ref{app}. We should point out the differencing scheme derived by the {\it 
finite volume method} (Equation [\ref{eq:Lr2}]) is different from those used 
in GALPROP \citep{galp} or DRAGON \citep{2008JCAP...10..018E}; the numerical 
schemes of GALPROP and DRAGON are basically applicable for continuously and 
slowly changed diffusion coefficient.

Then we describe the parameter settings. In the difference equation, the time 
step $\Delta t$ is set to be 1000 years, which is much smaller than the age of 
Geminga. The spatial step $\Delta r$ is 1 pc, then the initial condition given 
in Equation (\ref{eq:ini}) describes a spherical source with a radius of 1 pc. 
For an old source like Geminga, the spectrum at the Earth can be very close to 
that of a point source, especially when $E\gtrsim100$ GeV 
\citep{2012MNRAS.419..624T}. We use a logarithmic scale with a ratio of 1.2 for 
the energy grids, that is, $E_{l+1}/E_l=1.2$, where $l$ is the energy step 
index. The $e^\pm$ injection spectrum takes the form of $Q(E)=Q_0E^{-\gamma}$, 
where $\gamma\simeq2.2$, as indicated by the HAWC observation 
\citep{2017Sci...358..911A}. 

\section{Result}\label{sec:rst}

\begin{figure}[!htb]
 \centering
 \includegraphics[width=0.55\textwidth]{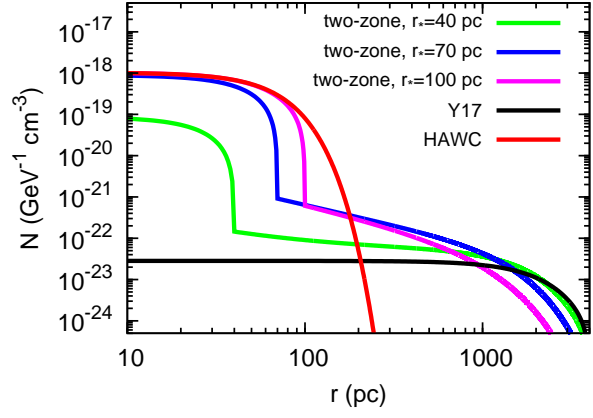}
 \caption{Radial distributions of 1 TeV $e^\pm$ from Geminga at the present 
age. Three different $r_\star$ are adopted for the two-zone diffusion models. 
The distributions of the one-zone diffusion cases are also shown for 
comparison, the black line is for the one-zone diffusion model with Y17 
diffusion coefficient, and the red line is calculated with parameters inferred 
by HAWC data \cite{2017Sci...358..911A}.}
 \label{fig:test}
\end{figure}

If $D$ is spatially uniform, Equation (\ref{eq:prop}) can be solved 
analytically using the Green's function method \citep{1964ocr..book.....G}. 
We verify that our numerical solution for the one-zone case matches well
with the analytical solution. For the two-zone diffusion case, there is no 
analytical way to test the results. To check that there is no ``swallowing'' or 
``spitting'' of particles in the discontinuity surface, we integrate the number 
of particles, $N(r_i)$, for the two-zone diffusion cases with different 
$r_\star$. We find that the number of particles is always the same as that of 
the one-zone cases. The present radial distributions of 1 TeV $e^\pm$ 
are shown in Figure \ref{fig:test} for the two-zone diffusion scenarios and 
also the one-zone cases. Here we assume that all the spin-down energy of 
Geminga pulsar is converted to the energy of injected $e^\pm$ (conversion 
efficiency), to determine the normalization.

\begin{figure}[!htb]
 \centering
 \includegraphics[width=0.48\textwidth]{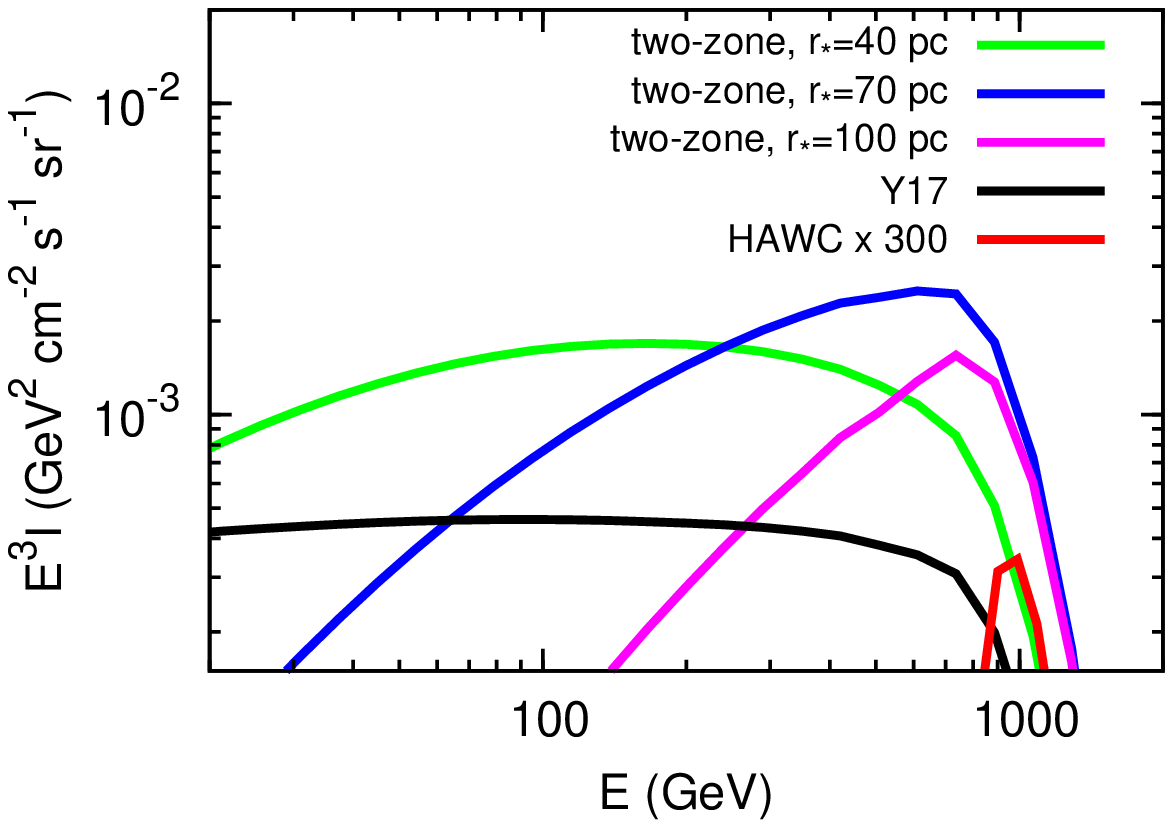}
 \includegraphics[width=0.48\textwidth]{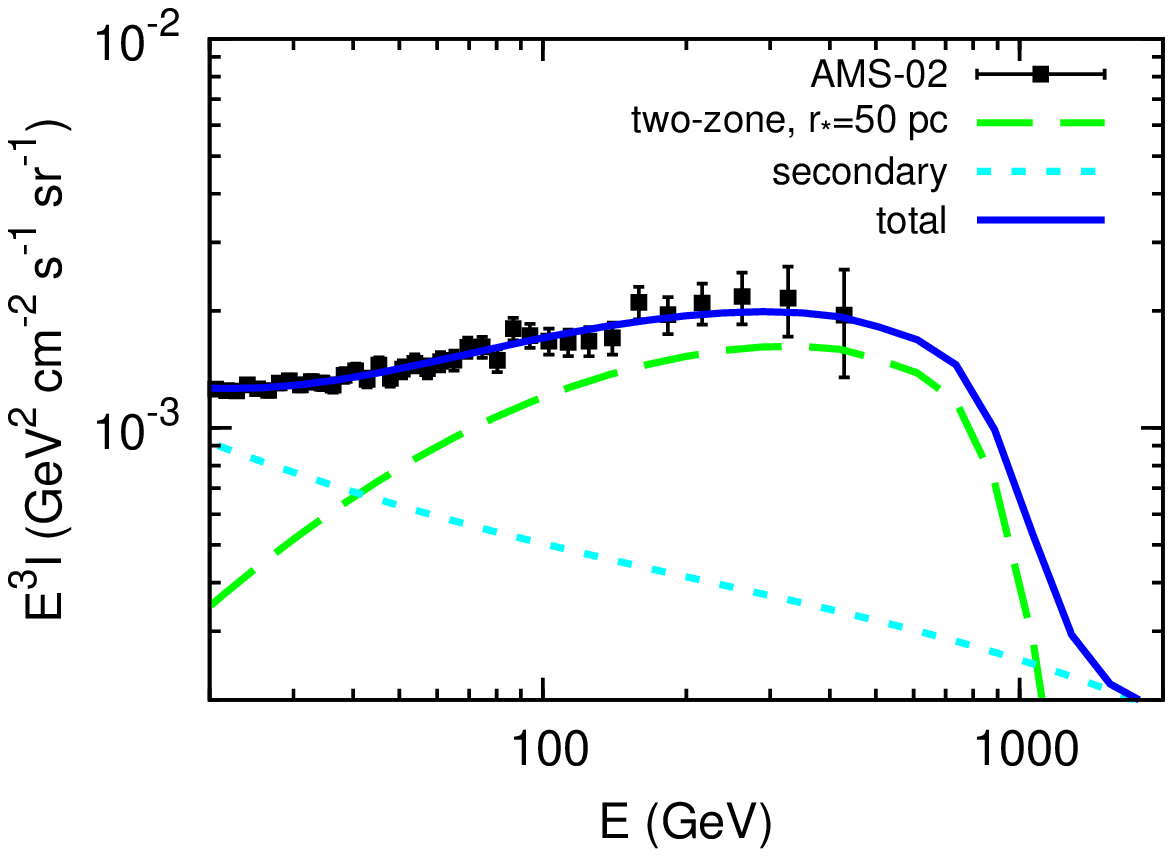}
 \caption{Left: expected electron (or positron) spectra of Geminga at the 
Earth. The conversion efficiency of Geminga pulsar is set to be 100\%. 
Three different values of $r_\star$ are adopted for the two-zone diffusion 
scenario. The spectra of the one-zone diffusion cases with $D_1$ (HAWC) and 
$D_2$ (Y17) are also presented. Note the former is 300 times of its original 
spectrum. Right: the model predicted positron spectrum compared with the AMS-02 
data \citep{2014PhRvL.113l1102A}. The green dashed line is the contribution 
of Geminga, with $r_\star=50$ pc and a conversion efficiency of 75\%.}
 \label{fig:spec}
\end{figure}

The positron (or electron) spectra at the Earth generated by Geminga are 
shown in the left panel of Figure \ref{fig:spec} for different $r_\star$. 
The conversion efficiency of Geminga pulsar is also set to be 100\%. 
The spectrum becomes harder for a larger $r_\star$, since fewer 
low energy $e^\pm$ can reach the Earth when the slow-diffusion zone is 
larger. The extreme cases of one-zone diffusion with Y17 or HAWC diffusion
coefficients are also shown for comparison. We find that at high energies, 
the fluxes of the two-zone diffusion models are considerably higher than 
that of the one-zone diffusion case with Y17 parameters. For a fast-diffusion 
model like Y17, the propagation scale of 1 TeV $e^\pm$ of Geminga is 
$\simeq1.7$ kpc. This scale is 
about 7 times larger than the distance between Geminga and the Earth, 
which implies that the particles have diffused to a considerably large 
region. For the two-zone diffusion scenario, the propagation of particles 
is hindered within the slow-diffusion zone. After being confined for
some time in the inner region, these particles enter the fast propagation
region and diffuse to the Earth. Since they spend less time in the
fast-diffusion region, they can be effectively accumulated in a relatively 
smaller region and result in a higher flux at the Earth's location.

We find that the two-zone model with $r_\star=50$ pc can well reproduce
the AMS-02 data of the positron flux. The results for $r_\star>50$ pc
are too hard to explain the data, while the spectrum is too soft for the case 
with a smaller $r_\star$ or the one-zone fast-diffusion model. Note that the 
final spectra depend on the injection spectrum of $e^\pm$ assumed, which is 
$\gamma \simeq 2.2$ in this work. The comparison between the model and the data 
is shown in the right panel of Figure \ref{fig:spec}. Here the conversion 
efficiency to $e^\pm$ is assumed to be 75\% in order to match the data. For the 
secondary contribution from $pp$ collisions, one can refer to \citet{dela09} 
and \citet{2018ApJ...854...57F}. 

\begin{figure}[!htb]
 \centering
 \includegraphics[width=0.55\textwidth]{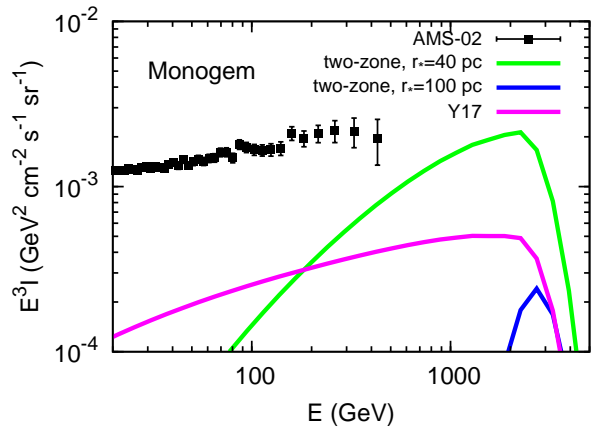}
 \caption{Positron spectra of Monogem with different diffusion models, 
compared with the AMS-02 data. The conversion efficiency is 100\% for all the 
cases.}
 \label{fig:spec_m}
\end{figure}

The $e^\pm$ spectrum of Monogem under the two-zone diffusion can be obtained in 
the same way. The distance, age, and spin-down energy of Monogem is 290 pc, 
$1.1\times10^5$ years, and $1.58\times10^{48}$ erg, respectively 
\citep{2005AJ....129.1993M}. The observation of HAWC indicates a spectral index 
of $\sim$ 2.0 for the injection spectrum of Monogem 
\citep{2017Sci...358..911A}. As can be seen from Figure \ref{fig:spec_m}, the 
positron spectra of Monogem is too hard to fit the AMS-02 data in the two-zone 
diffusion cases, due to its farther distance, younger age, and harder injection 
spectrum compared with Geminga. Thus Monogem cannot make a major contribution 
to the local positron fluxes even in the two-zone diffusion model.

\begin{figure}[!htb]
 \centering
 \includegraphics[width=0.55\textwidth]{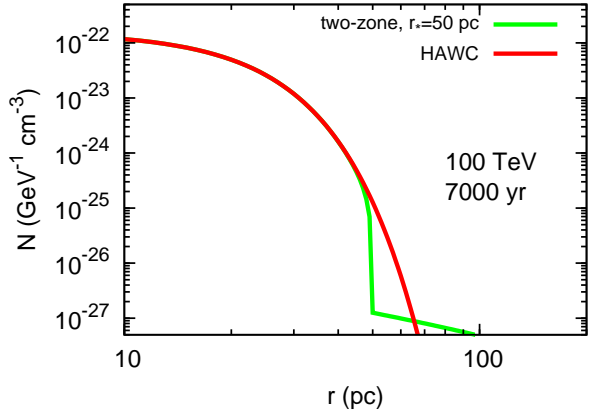}
 \caption{Density profiles of 100 TeV $e^\pm$ with diffusion of 7000 
years. The red line describes the slow-diffusion case proposed by 
\cite{2017Sci...358..911A}, while the green line represents the two-zone 
diffusion case with $r_\star=50$ pc.}
 \label{fig:comp}
\end{figure}

Finally we would like to give a simple self-consistency check if the 
two-zone region can reproduce the $\gamma$-ray profile observed by HAWC. 
The HAWC $\gamma$-ray data around 20 TeV are produced by $\sim100$ TeV
electrons, whose cooling time is about 7000 years. This indicates that the 
parent $e^\pm$ of the $\gamma$-ray observed by HAWC should be younger than 7000 
years. We show in Figure \ref{fig:comp} the density profiles of 100 TeV 
$e^\pm$ from Geminga with diffusion of 7000 years. Within 30 pc from Geminga, 
we find very good consistency between the one-zone slow-diffusion scenario of 
\cite{2017Sci...358..911A} and the two-zone diffusion scenario with 
$r_{\star}=50$ pc, which means the later proposed in this work can also 
accommodate the HAWC $\gamma$-ray data. The reason for this consistency is that 
7000 years is too short for $e^\pm$ to escape from the slow-diffusion region 
of the two-zone model.

\section{Conclusion and Discussion}
\label{sec:discuss}
In this work we propose that a two-zone propagation scenario of cosmic
ray $e^\pm$ to account for the HAWC observations of extended halo
around Geminga pulsar and the locally observed positron flux by AMS-02.
The diffusion of $e^\pm$ is assumed to be significantly slow within a 
distance of $r_\star$ from the pulsar, as inferred by the spatial
brightness profile of $\gamma$-ray emission observed by HAWC
\citep{2017Sci...358..911A}. The particles diffuse beyond
$r_\star$ with a diffusion coefficient inferred from the B/C data.
A numerical method is adopted to solve the propagation equation, and the 
differencing scheme of the diffusion operator is derived by the {\it finite 
volume method}, which is different from those used in GALPROP or DRAGON.

The exact value of $r_\star$ to fit the AMS-02 positron flux degenerates 
with the injection spectrum of $e^\pm$, and hence can neither be too big 
nor too small. For larger (smaller) value of $r_\star$, low energy 
particles would be confined in the slow-diffusion region longer 
(shorter), and the final positron spectrum is harder (softer).
For an injection spectral index of $\sim 2.2$ and $r_\star \sim 50$ pc, 
we find that the $e^\pm$ from Geminga can reasonably account for the 
positron excess, with about 75\% of the spin-down energy converted
into $e^\pm$. Actually such a result is even more natural than that 
in the one-zone diffusion scenario which requires an efficiency even
slightly larger than 100\% \citep{2013PhRvD..88b3001Y}. 

The origin of the slow-diffusion zone around Geminga is still unclear. It is 
possible that the zone is pre-existed. For example, the shock of the parent 
supernova remnant (SNR) of Geminga, which is non-observable today, may have 
swept the ISM and made it more turbulent. Considering the age of $3\times10^5$ 
years, the shocked region can reach a size of $\sim$100 pc \citep{yamazaki06}. 
We note that Geminga has a transverse velocity of $205^{+90}_{-47}$ km s$^{-1}$ 
\citep{2007Ap&SS.308..225F}, which means a 70 pc offset from its birth place. So 
Geminga may still be within the shocked region of its SNR, where the diffusion 
coefficient is smaller. Meanwhile, if most of $e^\pm$ are injected in 
the early age of Geminga, when the offset of Geminga is small and the scale of 
its SNR is not so large as today, our assumption of the symmetrical 
slow-diffusion zone may not be impacted.

Alternatively, the slow diffusion zone can be generated intrinsically by the 
$e^\pm$ injected from Geminga. Near the cosmic-ray sources, the spatial gradient 
of particle density is significantly larger than that of the average ISM, which 
can lead to the growth of the streaming instability 
\citep{2008AdSpR..42..486P,malkov13,blasi16}. In this case, the 
magnetohydrodynamic turbulence may be considerably stronger than that in the 
ISM. Therefore, it is plausible that the $e^\pm$ leaving Geminga are confined in 
the nearby zone for a longer time by the waves induced by themselves. 

The slow-diffusion regions around pulsars may be common, which may result
in extended $\gamma$-ray halos of pulsars and can probably explain the
diffuse $\gamma$-ray excess observed in the Galactic plane 
\citep{2008ApJ...688.1078A,2017arXiv170701905L,2016ChPhC..40k5001G}.

\acknowledgments{This work is supported by the National Key Program for Research 
and Development (No.~2016YFA0400200) and by the National Natural Science 
Foundation of China under Grants No.~U1738209,~11475189,~11475191,~11722328. Q. 
Yuan acknowledges the 100 Talents Program of Chinese Academy of Science.}

\appendix

\section{Discretization of the Propagation Equation}
\label{app}
We adopt the \emph{operator splitting method} to divide the propagation 
equation into two one-dimensional sub-problems, and choose the \emph{Strang 
splitting} which is second-order accurate \citep{strang}. We depict the flow 
chart as
$$N^{n}\xrightarrow[(t_n,\,t_{n+1/2})]{\mathcal{L}_E}\tilde{N}^{n+1/2}
\xrightarrow 
[(t_n,\,t_{n+1})]{\mathcal{L}_r}N^{n+1/2}\xrightarrow[(t_{n+1/2},\,t_{n+1})]{ 
\mathcal{L}_E}N^{n+1}\,,$$
where $n$ is the time step index. \citet{2014APh....55...37K} uses trapezoidal 
integration for the discretization of the energy-loss scheme, which leads to a 
second-order accuracy. The algebraic equation is written as
\begin{equation}
 \begin{aligned}
   N^{n+1}_{l,i}= & \left(\frac{b_{l+1}\Delta t-\Delta E}{b_{l}\Delta 
t+\Delta E}\right)N^{n+1}_{l+1,i} - \left(\frac{b_{l}\Delta t-\Delta 
E}{b_{l}\Delta t+\Delta 
E}\right)N^{n}_{l,i} \\
  & + \left(\frac{b_{l+1}\Delta t+\Delta E}{b_{l}\Delta t+\Delta 
E}\right)N^{n}_{l+1,i}\,\,, \\
 \end{aligned}
 \label{eq:LE}
\end{equation}
where $l$ is the energy step index, $i$ is the spatial step index, $\Delta t$ 
denotes the time step, and $\Delta E=E_{l+1}-E_l$. 

For the diffusion operator, we use two different discretizations, depending on 
if $r_i$ lies on the discontinuity surface. If not, we apply the Crank-Nicolson 
scheme for constant $D$, which is second-order accurate and unconditionally 
stable. The corresponding algebraic equation is
\begin{equation}
 \begin{aligned}
  & -\frac{D_{l,i}\Delta t}{2\Delta r^2}\left(1-\frac{\Delta r}{r_i}
 \right)N^{n+1}_{l,i-1} + \left(1+\frac{D_{l,i}\Delta t}{\Delta 
r^2}\right)N^{n+1}_{l,i} - \frac{D_{l,i}\Delta 
t}{2\Delta r^2}\left(1+\frac{\Delta r}{r_i}\right)N^{n+1}_{l,i+1} \\
  & = \frac{D_{l,i}\Delta t}{2\Delta r^2}\left(1-\frac{\Delta r}{r_i}
 \right)N^{n}_{l,i-1} + \left(1-\frac{D_{l,i}\Delta t}{\Delta 
r^2}\right)N^{n}_{l,i} + \frac{D_{l,i}\Delta 
t}{2\Delta r^2}\left(1+\frac{\Delta r}{r_i}\right)N^{n}_{l,i+1}\,,\\
 \end{aligned}
 \label{eq:Lr1}
\end{equation}
where $\Delta r$ is the radial step size, and $r_i=i\Delta r$. Equation 
(\ref{eq:Lr1}) is a tridiagonal system, which can be solved with \emph{LU} 
decomposition. For the case of $r_i=r_\star$, the difference 
equation should be re-written to ensure the conservation of the flux at 
both sides of the discontinuity surface. We adopt the \emph{finite volume 
method} to derive the differencing scheme for this case, which reads
\begin{equation}
 \begin{aligned}
  & - \frac{D_{l,i-1}\Delta t}{2\Delta r^2}\left(1-\frac{\Delta 
r}{r_i}\right)N^{n+1}_{l,i-1} \\
  & + \left[1+\frac{D_{l,i-1}\Delta t}{2\Delta 
r^2}\left(1-\frac{\Delta 
r}{r_i}\right)+\frac{D_{l,i+1}\Delta t}{2\Delta r^2}\left(1+\frac{\Delta 
r}{r_i}\right)\right]N^{n+1}_{l,i} \\
  & -\frac{D_{l,i+1}\Delta t}{2\Delta r^2}\left(1+\frac{\Delta 
r}{r_i}\right)N^{n+1}_{l,i+1} \\
  & = \frac{D_{l,i-1}\Delta t}{2\Delta 
r^2}\left(1-\frac{\Delta 
r}{r_i}\right)N^{n}_{l,i-1} \\
  &\quad + \left[1-\frac{D_{l,i-1}\Delta t}{2\Delta 
r^2}\left(1-\frac{\Delta 
r}{r_i}\right)-\frac{D_{l,i+1}\Delta t}{2\Delta r^2}\left(1+\frac{\Delta 
r}{r_i}\right)\right]N^{n}_{l,i} \\
  &\quad +\frac{D_{l,i+1}\Delta t}{2\Delta r^2}\left(1+\frac{\Delta 
r}{r_i}\right)N^{n}_{l,i+1} \,\,.\\
 \end{aligned}
 \label{eq:Lr2}
\end{equation}
The derivation of this scheme is presented as follows.

Consider a diffusion equation
\begin{equation}
 \frac{\partial N}{\partial t}-\frac{1}{r^2}\frac{\partial}{\partial 
r}\left[r^2D(r)\frac{\partial N}{\partial r}\right]=0\,,
 \label{eq:app1}
\end{equation}
where $D(r)$ is defined in Equation (\ref{eq:diff}). We set 
$F=r^2D(r)\partial N/\partial r$, which is the particle flux in $r$. Then 
Equation (\ref{eq:app1}) can be rewritten as 
\begin{equation}
 \frac{\partial Nr^2}{\partial t} = \frac{\partial F}{\partial r}\,.
 \label{eq:app2}
\end{equation}
Assuming $r_i=r_\star$, the continuity of particle density and flux should be 
satisfied:
\begin{equation}
 \left\{
 \begin{aligned}
  & N(t, r_i-0) = N(t, r_i+0)\,, \\
  & F(t, r_i-0) = F(t, r_i+0)\,. \\
 \end{aligned}
 \right.
 \label{eq:app3}
\end{equation}
We integrate Equation (\ref{eq:app2}) in the time range of 
$[t_n, t_{n+1}]$ and the radial range of $[r_{i-1/2}, r_{i+1/2}]$, where 
$r_{i\pm1/2}=r_i\pm\Delta r/2$, and we obatin
\begin{equation}
 \begin{aligned}
  & \int_{r_{i-1/2}}^{r_{i+1/2}}[N(t_{n+1}, r)-N(t_n, 
r)]r^2dr \\
  & 
=\int_{t_n}^{t_{n+1}}\left[\int_{r_{i-1/2}}^{r_{i+1/2}}\frac{ 
\partial}{\partial r}F(t, r)dr\right]dt \\
  & =\int_{t_n}^{t_{n+1}}\left[\int_{r_{i-1/2}}^{r_{i}}\frac{ 
\partial}{\partial r}F(t, r)dr + \int_{r_{i}}^{r_{i+1/2}}\frac{ 
\partial}{\partial r}F(t, r)dr\right]dt \\
  & =\int_{t_n}^{t_{n+1}}\{[F(t, r_i-0)-F(t, r_{i-1/2})]+[F(t, 
r_{i+1/2})-F(t,r_i+0)]\}dt \\
  & =\int_{t_n}^{t_{n+1}}[F(t, r_{i+1/2})-F(t,r_{i-1/2})]dt\,. 
\\
 \end{aligned}
 \label{eq:app4}
\end{equation}
On the other hand, we have $\partial N/\partial r=F/[r^2D(r)]$ as defined 
above. Integrating this relation in $[r_i, r_{i+1}]$, we get
\begin{equation}
 \begin{aligned}
  N(t, r_{i+1})-N(t, r_i) &= \int_{r_i}^{r_{i+1}}\frac{F(t,r)}{r^2D(r)}dr \\
                                    &\approx F(t, 
r_{i+1/2})\int_{r_i}^{r_{i+1}}\frac{1}{r^2D(r)}dr\,,\\
 \end{aligned}
 \label{eq:app5}
\end{equation}
which means
\begin{equation}
 F(t,r_{i+1/2})\approx\frac{N(t, r_{i+1})-N(t, 
r_i)}{\displaystyle\int_{r_i}^{r_{i+1}}\dfrac{1}{r^2D(r)}dr}\,.
 \label{eq:app6}
\end{equation}
The expression of $F(t,r_{i-1/2})$ can be found similarly. The integral terms 
in Equation (\ref{eq:app4}) can be approximated by
\begin{equation}
 \int_{r_{i-1/2}}^{r_{i+1/2}}N(t_n, r)r^2dr \approx N(t_n, r_i)r^2_i\Delta r
 \label{eq:app7}
\end{equation}
and
\begin{equation}
 \int_{t_n}^{t_{n+1}}F(t, r_{i+1/2})dt \approx [F(t_{n+1}, 
r_{i+1/2})+F(t_n, r_{i+1/2})]\Delta t/2\,,
 \label{eq:app8}
\end{equation}
where the latter corresponds to the Crank-Nicolson scheme. Combining Equation 
(\ref{eq:app4}) and (\ref{eq:app6}--\ref{eq:app8}), we finally obtain
\begin{equation}
 \begin{aligned}
  & \frac{N(t_{n+1}, r_i)-N(t_n, r_i)}{\Delta t} = \\
  & \frac{D_{i+1}r_{i+1}}{2\Delta r^2r_i}[N(t_{n+1}, r_{i+1})-N(t_{n+1}, 
r_i)+N(t_{n}, r_{i+1})-N(t_{n}, r_i)] \\
  & - \frac{D_{i-1}r_{i-1}}{2\Delta r^2r_i}[N(t_{n+1}, r_{i})-N(t_{n+1}, 
r_{i-1})+N(t_{n}, r_{i})-N(t_{n}, r_{i-1})]\,, \\
 \end{aligned}
 \label{eq:app9}
\end{equation}
which can be rewritten as Equation (\ref{eq:Lr2}).

Finally, the initial condition is approximated by a step function
\begin{equation}
 N(0, E, r_i)=\left\{
 \begin{aligned}
  & Q'(E),  & i\leq1 \\
  & 0,  & i>1 \\
 \end{aligned}
 \right. \,,
 \label{eq:ini}
\end{equation}
where $Q'(E)=Q(E)/(4\pi r^3_1)$. For the inner spatial boundary condition, 
we handle it with $N(r_1)=N(r_{-1})$. Plugging this relation into Equation 
(\ref{eq:Lr1}), we can then obtain $N(r_0)$.

\bibliography{geminga}

\end{document}